\documentclass[11pt,3p]{elsarticle}

\makeatletter
\def\ps@pprintTitle{%
 \def\@oddfoot{\centerline{\thepage}}%
 \let\@evenfoot\@oddfoot}
\makeatother

\usepackage{amssymb}
\usepackage{graphicx} % Include figure files
\usepackage{dcolumn}  % Align table columns on decimal point
\usepackage{bm}       % bold math
\usepackage{color}
\usepackage{mathrsfs}
\usepackage{nicefrac}
\usepackage{xcolor,colortbl}

\usepackage{url}
\usepackage[hidelinks,colorlinks=true,allcolors={blue},breaklinks]{hyperref}
\usepackage[caption=false]{subfig}
\begin{document}

\begin{frontmatter}

%\title{Numerical study of self-avoiding walking properties from a process of finite time growth without border limits in a cubic lattice: a reasonable model for polymeric linear chains.}

\title{A simple self-avoiding walking process as a reasonable non-conventional generator of polymeric linear chains.}

\author[label1]{David R. Avellaneda B.\corref{cor1}}
\address[label1]{Departamento de Estat\'{i}stica e Inform\'{a}tica, Universidade Federal Rural de Pernambuco, Recife, Pernambuco, CEP 52171-900, Brazil}

\cortext[cor1]{Corresponding author.}
\ead{davidave16@gmail.com}

\author[label2]{Ram\'{o}n E. R. Gonz\'{a}lez}
\address[label2]{Departamento de F\'{i}sica, Universidade Federal Rural de Pernambuco, Recife, Pernambuco, CEP 52171-900, Brazil.}

\ead{ramon.ramayo@ufrpe.br}

\begin{abstract}
In this work, we present a simple and efficient generator of polymeric linear chains, based on a  random self-avoiding walk process. The chains are generated using a discrete process of growth, in cubic networks and in a finite time, without border limits and without exploring all the configurational space. First, we thoroughly describe the chains morphology exploring the statistics of two characteristic distances, the radius of gyration and the end-to-end distance. Moreover, we examine the dependence of mean characteristic distances with the number of steps ($N$). Despite the simplicity of our procedure, we obtain universal critical exponents, which are in very good agreement with previous values reported in the literature. Moreover, studying the balance between the monomer-monomer interaction and the bending energy, we find that initially, the chains develop by multiple doubling, forming a cluster and increasing its energy. After reaching a given number of steps, the chains stretch and flee from the cluster, which results in a reduction of its interaction energy. However, the behaviour of the bending energy reveals that the chains follow the same folding pathway in both regimes. Additionally, we also characterize the energy of the obtained chains, combining the local interaction energy with its corresponding bending energy but in a discrete version. This analysis is relevant because it allows differentiating between chains of equal interaction energy but with different structures.
\end{abstract}

\begin{keyword}
%% keywords here, in the form: keyword \sep keyword
Self-avoiding random walk \sep polymer chains \sep critical exponents \sep interaction and bending energy \sep radius of gyration \sep end-to-end distance.
%% MSC codes here, in the form: \MSC code \sep code
%% or \MSC[2008] code \sep code (2000 is the default)
\end{keyword}

\end{frontmatter}

%========================================================================
%========================================================================
\section{INTRODUCTION}\label{Intro}

Although the conformational properties of polymer chains in a good solvent have been the subject of intensive experimental, theoretical and numerical studies their full understanding is still an open challenge. 
Besides its simplicity, the natural self-avoiding random walk (SAW), proposed over half a century ago, is widely accepted as the principal model for dilute polymers \cite{book_flory, medio_seculo1, cn2d}. 
SAW describes well a large spectrum of real systems with diverse details (such as bond angles and monomer-monomer potential).
Moreover, the equivalence of SAW with the $n\rightarrow 0$ limit of the $n$-vector model \cite{nvector} has provided an important connection with the theory of phase transitions and critical phenomena \cite{degennes}.\\

One of the first (but still widely used) theoretical approaches to this subject is the Flory theory \cite{book_flory}, which with simple mean field arguments involving the concept of excluded volume brought about the understanding of underlying power laws and the role of dimensionality.
The subsequent analytical approaches span rigorous methods that have achieved only limited success \cite{SAW1993}, approximate methods such as perturbation theory and self-consistent field theory, which break down for long chains \cite{yamakawa1971}, and renormalization group (RG) \cite{WILSON1974} which has yielded reasonably accurate estimates for critical exponents and some universal amplitude ratios.\\

In parallel with theoretical developments, numerical methods have shown to be a fundamental tool in establishing properties of long SAWs \cite{sokal1995}. 
Exact enumeration methods have been used to find the number of all possible SAWs of finite length $N$, from which universal properties are estimated using techniques such as the ratio method, Pade approximants or differential approximants. Thus, results up to $N=71$ steps for the square lattice \cite{guttmann2001,jensen2004} and up to $N=21$ steps for the cubic lattice \cite{SAW1993} have been reported. A large number of Monte Carlo sampling techniques have been proposed ever since the 1950s (see, e.g., \cite{sokal1995} for a comprehensive overview), where the pivot algorithm \cite{medio_seculo1} has been shown to perform in time $O(N)$, and studies with SAWs of length up to $N=80000$ have been reported \cite{Li1995}.\\

Typically, an ensemble of self-repelling chains considers all possible configurations of a given length.
Moreover, the growth process is directed by probability values, which are linked to the Boltzmann factor in terms of the potential energy of interaction and are proportional to the number of interactions \cite{trueSAW_Peliti}.\\

Rather than attempting to improve the performance previous algorithms, in the current work we focus on different goals.
Instead, in this paper we study relatively small SAW chain ensembles, which might be relevant describing polymer solutions where the polymer growth process has been going on for a limited time.
We present a method to generate such ensembles using very simple and efficient numerical algorithm and, surprisingly, we obtain universal critical exponents, which are in very good agreement with the previous reported in the literature. 
In addition, we carefully examine the balance between the monomer-monomer interaction and the bending energy, addressing its relation with the structure of the chains. This analysis is relevant because it allows differentiating between chains of equal interaction energy but with different structures.\\

The paper is organized as follows. In the next section we briefly review the SAW and the principal conformation measures, the algorithm, and the energy measures. 
The subsequent section is devoted to the results of our simulations, and finally the conclusions are drawn.

%========================================================================
%========================================================================
\section{METODOLOGY} \label{Metodologia}

Our aim is to efficiently generates an ensemble of linear homopolymer chains in a good solvent.
For simplicity, we use the approximation that the solvent molecules are considered the same size as the monomers. \\

The chains are generated using the pathway of a particle that moves randomly in a cubic network with an unlimited boundary conditions. Following that approach, the particle is not allowed to occupy the sites it has visited before (a particle with property self-avoiding). We use the idea of a model known as  ``true'' self-avoiding random walk \cite{trueSAW_Peliti} but in 3d. Thus, each steps given by the walker can be interpreted as a monomer (or a set of monomers of the same type), and the steps not visited by the walker (empty sites of the network) can be considered as molecules of the solvent, thus, the trajectory described by this particle defines a homopolymeric chain in a good solvent \cite{trueSAW_Peliti}. 
The probability of the walker taking a step in the direction $i$ in this type of chain depends on the number of times $n_j$ that the next time the site to be occupied is already ``visited'', and is given by:
%%%%%%%%%%%%%%%%
\begin{equation}
p_i = \frac{e^{-gn_{i}}}{\displaystyle\sum_{j=1}^{3d}e^{-gn_{j}}}= \frac{1}{\displaystyle\sum_{j=1}^{3d}e^{-g(n_{j}-n_{i})}}.
\label{eq:trueSAW}
\end{equation}
%%%%%%%%%%%%%%%%

In general, the sum runs through all possible $3d$ paths from the position occupied by the walker at each instant of time, including the address $i$, and $g$ is a positive parameter which measures the intensity with which the walk avoids it self. 
For the sake of simplicity, in this work we implement the limiting case $g=\infty$, which corresponds in the $3d$ case to a discrete domain of probabilities $p_i(g=\infty)=[1/6, 1/5, 1/4, 1/3, 1/2, 1]$.

%========================================================================
\subsection{Numerical Algorithm} \label{Algoritmo}

For a $d-$dimensional network with free boundaries the algorithm to generate a one chain is as follows:

%%%%%%%%%%%%%%%%
\begin{enumerate}%[\bf i.]
	\item Choose the number of attempts $N'$.
	\item Choose the origin of the polymer, which in our case is the origin of the coordinate system.
	\item Generate the first step randomly or choose it arbitrarily from a point in the cubic network.
	\item Choose the following step randomly from one of the $2\times d$ possible steps.
	\item If the given step leads to self-intersection, go to item 4. and try again with another step. This step is most important to ensure the SAW.
	\item If the step leads to an available location, add the step to the walk.
	\item If the number of attempts is reached or if the number of possible steps is zero (the walker gets stuck), the simulation is accepted and saved. 
\end{enumerate}
%%%%%%%%%%%%%%%%

Thus, the random chain is formed by $N$ steps generated from $N'$ attempts, being that, for the three-dimensional case $N'>N$, this is because the chains can get trapped before reaching the total number of attempts.\\

After generating the random chain, we store the positions of each of the monomers that constitute it and proceed to compute the characteristic measurements of its configuration. 
Starting by calculating the displacement ${\bf{r}}_i$ of each monomer in order to obtain the mass center ${\bf{r}}_c$ given in Eq. (\ref{eq:centro_de_massa}) and with which we calculate the radius of gyration, ${\bf{R}}_g$, given in Eq. (\ref{eq:raio_de_giro}). 
Next, we compute the end-to-end vector module, ${\bf{R}}_{ee}$, which can be easily derived from the distance of the $N$th monomer to the origin of the chain as described in Eq. (\ref{eq:Ree}).\\

%========================================================================

\subsection{Characteristic distances}

As is proven below, this simple algorithm efficiently generates an ensemble of linear homopolymer chains in a good solvent; each of them  is formed by $N$ monomers in positions $\{r_{0},r_{1},...,r_{N}\}$ in the space of dimension $d$. The separation distance between a monomer and its nearest neighbor is $b = r_{i} - r_{i-1}$, for $i=1,2,...,N$, which would be equivalant to a Kuhn segment \cite{book_flory,Rubinstein}.\\

Moreover, we thoroughly describe the chains morphology exploring the behaviour of two characteristic distances, the end-to-end distances ${\bf{R}}_{ee}$ and the radius of gyration ${\bf{R}}_g$.
The end-to-end distance is defined as the mean squared variance of the displacement and is reads as,
%%%%%%%%%%%%%%%%
\begin{equation}
{\bf{R}}_{ee}^{2} = \langle({\bf{r}}_N-{\bf{r}}_0)^2\rangle,
\label{eq:Ree}
\end{equation}
%%%%%%%%%%%%%%%%
the variables ${\bf{r}}_N$ and ${\bf{r}}_0$ are the positions of the ends of the chain. 
The radius of gyration ${\bf{R}}_g$, whose square is the second moment around the center of mass ${\bf{r}}_c$ given by:
%%%%%%%%%%%%%%%%
\begin{equation}
{\bf{r}}_c = \frac{1}{N+1} \sum_{i=0}^N {\bf{r}}_i,
\label{eq:centro_de_massa}
\end{equation}
%%%%%%%%%%%%%%%%
%%%%%%%%%%%%%%%%
so that, the radius of gyration takes the form \cite{polimer_textbook,Rubinstein}:
%%%%%%%%%%%%%%%%
\begin{equation}
{\bf{R}}_g^2 = \frac{1}{N+1} \sum_{i=0}^N \langle({\bf{r}}_i-{\bf{r}}_c)^2\rangle.
\label{eq:raio_de_giro}
\end{equation}
%%%%%%%%%%%%%%%%

For real (non-Gaussian, with excluded volume) chains, a relationship between these distances is \cite{polimer_textbook}:
%%%%%%%%%%%%%%%%
\begin{equation}
\frac{6{\bf{R}}_g^2}{{\bf{R}}_{ee}^{2}} = 0.952.
\label{eq:relationship}
\end{equation}
%%%%%%%%%%%%%%%

For this type of chain, the radius of gyration depends on $N$, size of the Kuhn segment $b$ and exponent $\nu$:
%%%%%%%%%%%%%%%%
\begin{equation}
{\bf{R}}_g = bN^{\nu}.
\label{eq:exp_flory}
\end{equation}
%%%%%%%%%%%%%%%%
The exponent $\nu$ is called the size exponent or the Flory exponent \cite{book_flory,Rubinstein,flory_theor_polym}, which comes from Flory's theory, which had remarkable success in explaining the experimental evidence in the swelling of real polymers.\\

%========================================================================
\subsection{Energy of linear chains}\label{Energia}

In our analysis, we use a definition of the interaction energy, which is based on the compactness of the chains \cite{pedro,entropia_energia,energia_chao_tang}.
Complementary, we also examine the bending energy of the chains, which in our case is a discrete variable. 
This magnitude characterises the flexibility of the chains, as well as, its tangential correlations. Furthermore, the total energy of the chain is the sum of the bending energy and the interaction energy.

%========================================================================
\subsubsection{Interaction Energy Term}\label{energia_interacao}

\begin{figure}[h!]
\begin{center}
\includegraphics[width=6.5cm]{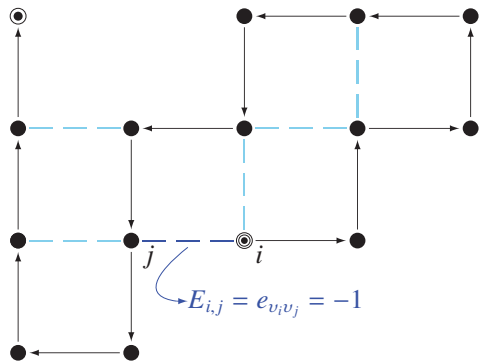}
\caption{14-step 2d-chain that shows the values that the interaction takes $\Delta({\bf{r}}_i-{\bf{r}}_j)$, where it adopts the values: $1$, for the attachments but not adjacent sites (blue color) and $0$ for the otherwise.}
\label{figura1}
\end{center}
\end{figure}

Interaction energy accounts for the energy of the chain due to its compactness and it quantifies the short-range interactions (Von Neumann neighborhood) for non-continuous monomers \cite{pedro,entropia_energia,energia_chao_tang,REM}. 
It reads as,
%%%%%%%%%%%%%%%%
\begin{equation}
E=\sum_{i<j}e_{\upsilon_{i}\upsilon_{j}}\Delta({\bf{r}}_i-{\bf{r}}_j),
\label{eq:energia_interacao}
\end{equation}
%%%%%%%%%%%%%%%%
where $\Delta({\bf{r}}_i-{\bf{r}}_j)=1$ if ${\bf{r}}_i$ and ${\bf{r}}_j$ are attached to the network, but $i$ and $j$ are not adjacent positions along the chain sequence and $\Delta({\bf{r}}_i-{\bf{r}}_j)=0$ for otherwise (see Fig. \ref{figura1}). 
The value of the factor, $e_{\upsilon_{i}\upsilon_{j}}$, depends on the type of contact between the monomers and represents the potential energy of interaction between the monomers located in the position ${\bf{r}}_i$ and ${\bf{r}}_j$ respectively. In our case, it is $e_{\upsilon_{i}\upsilon_{j}}= -1$ because we consider an attractive monomer-monomer interaction in a homopolymeric linear chain.

%%%%%%%
%%%%%%%
%========================================================================
\subsubsection{Bending Energy}\label{energia_bending}

One of the basic characteristics of all macromolecules is their flexibility \cite{Grosberg_solvent}. 
The polymer chains, in the pure state or in a dissolution, may adopt different conformations depending on their flexibility. When the flexibility is high, the chain may have large changes of direction within a few links. 
On the contrary, if the flexibility is low, the chain will be more rigid and will tend, in the limit, to behave as a hard stick. The flexibility of the polymer chain is related to the persistence length $l_p$. This can be defined as the average value of the maximum linear length of the chain configuration (it is also related to the Kuhn segment as $b=2l_{p}$ \cite{Rubinstein}). 
At distances greater than $l_p$, fluctuations in relation to itself or to the surroundings destroys the memory bound to the direction of the chain. Thus, the polymers are not completely flexible, some energy is required to fold them, which can happen up to at most the $l_p$.\\

\begin{figure}[h!]
\begin{center}
\centering
\subfloat[]{
\includegraphics[width=5.5cm]{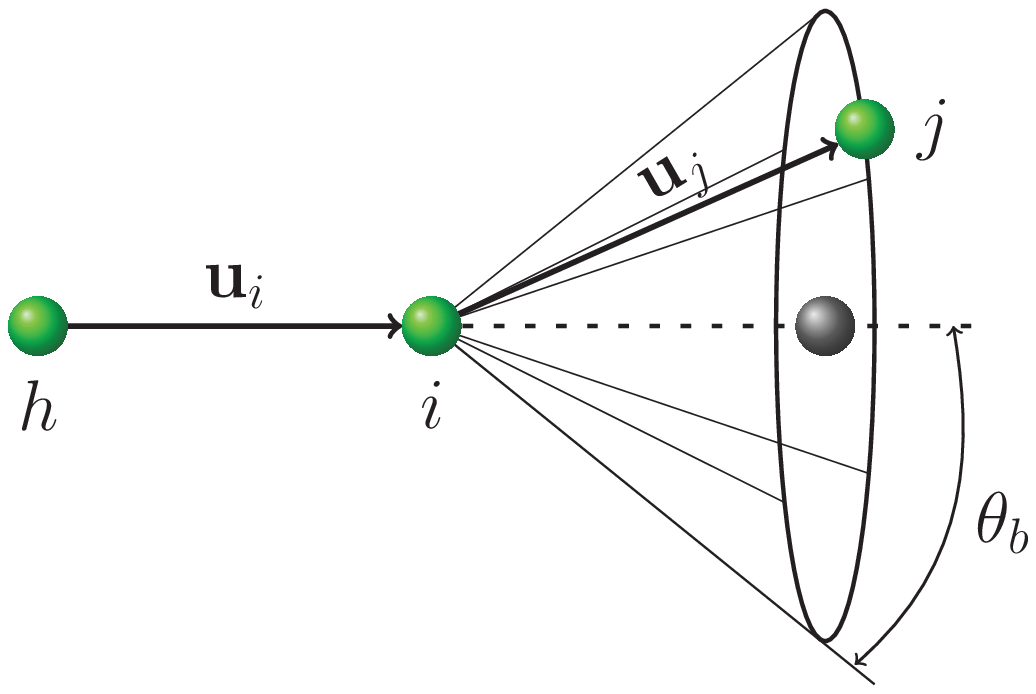}
\label{fig:tikz1}
}\hspace{1.0truecm}
\subfloat[]{
\includegraphics[width=4.5cm]{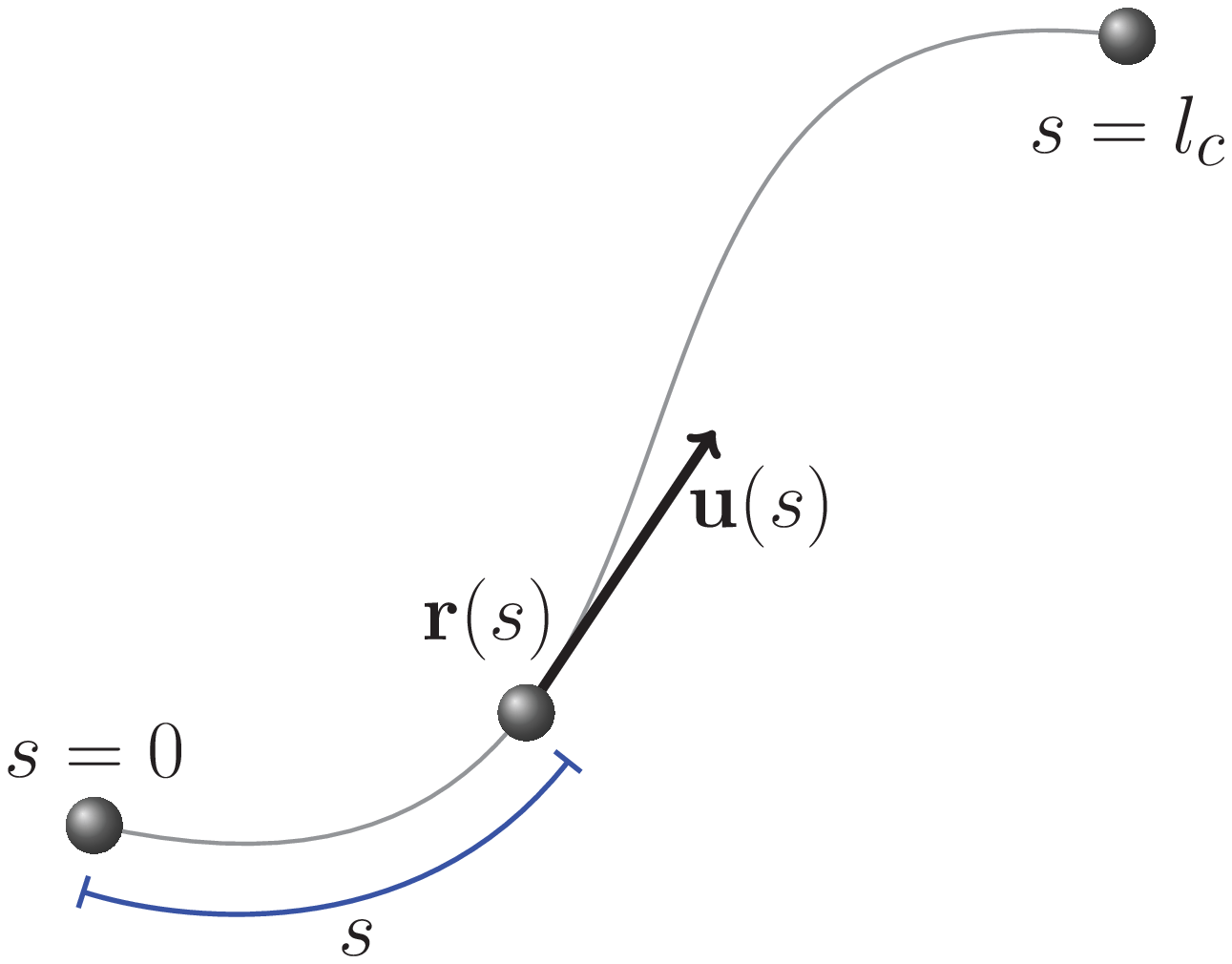}
\label{fig:tikz2}
}
\caption{Schematic representation of a section of the chain that shows high flexibility. (a) Vectors of bond ${\bf{u}}_i$ with a fixed link angle between two consecutive monomers. (b) Chain conformation specifying ${\bf{r}}(s)$ and the unit vector ${\bf{u}}(s)$.}
\label{fig:model_energy}
\end{center}
\end{figure}

The correlation (in turn related to the flexibility) between $\bf{u}$ and $\bf{u'}$, two unit vectors that join three points of the chain (monomers in the Fig. \ref{fig:tikz1}) and that are separated by a distance $l$ is given by \cite{polimer_textbook}:
%%%%%%%%%%%%%%%%%
\begin{equation}
\langle {\bf{u}} \cdot {\bf{u}}' \rangle = [(1-(b/l_{p}))^{1/b}]^{l} \sim \exp{(-l/l_{p})}.
\label{eq:correla_orient_exp}
\end{equation}
%%%%%%%%%%%%%%%%

%%%%%%%%%%%%%%%%

At the limit, the conformation of the chain is a smooth curve as described in Fig. \ref{fig:tikz2}. Using Eq. (\ref{eq:correla_orient_exp}) can obtain the correlation function between ${\bf{u}}(s)$ and ${\bf{u}}(s')$ of two segments of the chain, $s$ and $s'$, as a function of the persistence length $l_{p}$, given by \cite{polimer_textbook}:
%%%%%%%%%%%%%%%%
\begin{equation}
c(s,s')=\langle {\bf{u}}(s) \cdot {\bf{u}}(s')\rangle =  \exp{(-|s-s'|/l_{p})},
\label{eq:correlation}
\end{equation}
%%%%%%%%%%%%%%%%
this shows that the directional correlation of two segments of a macromolecule, decreases exponentially with the growth of the chain length \cite{Grosberg_solvent,flexibility}.\\

For bending energy we propose the following: we consider our polymer chain taking into account the interactions with other monomers of the same chain and the interactions with its environment. 
These interactions are described by an effective potential that represent the energy cost for its formation and whose stability is determined by two forces, one elastic with negative signal, that leads the chain to a collapse, and another repulsive of positive signal, that makes the chain is stretched. This energy cost is reflected in the chain in the form of free energy, for example, the number of conformations decreases with the increase of the vector end-to-end but increases the free energy of the same due to the high correlation that exists in the chain.\\

Chains with high flexibility experience changes of direction at a distance of few links tending to turn to itself, while low flexibility chains tend to become rigid, because the two-segment correlation function in the chain decreases exponentially with the distance between them as shown in Eq. (\ref{eq:correlation}). 
This correlated behavior occurs in the same way in a system of continuous mechanics, for the flexion model due to a force acting on a thin rod with stiffness constant $k$ \cite{Landau,energia_corr}. Flexing generates a differentiable curve on the rod, where, at a point ${\bf{r}}(s)$ of the curve there is a tangent vector ${\bf{u}}(s)$ generating a behavior similar to that described in Fig. \ref{fig:tikz2} for polymers. The Hamiltonian describing the internal energy of the rod of length $l_{c}$ is given by,
%%%%%%%%%%%%%%%%
\begin{equation}
\mathscr{H}=\frac{k}{2}\int_{0}^{l_{c}}\left(\frac{\partial{\bf{u}}(s)}{\partial s}\right)^{2} ds.
\label{eq:hamiltoniano_vara}
\end{equation}
%%%%%%%%%%%%%%%%

\begin{figure}[h!]
\begin{center}
\includegraphics[width=7.5cm]{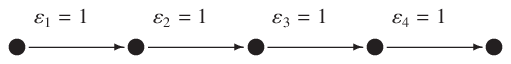}
\hspace{1.0cm}
\includegraphics[width=5cm]{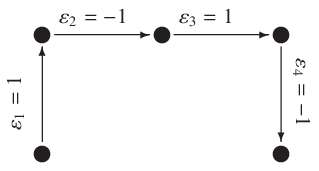}
\caption{4-step 2d-chain that shows the value of the weight function $\varepsilon_{i}$ used to compute the bending energy. Left: linear chain without deviation with its weight per step equal to $(1)$. Right: chain with mixed deviations, when direction changes $\varepsilon_{i}$ adopts a weight equal to $(-1)$, as is the case with steps $2$ and $4$.}
\label{figura3}
\end{center}
\end{figure}

By virtue of this, we propose a discretization of bending energy in order to be adapted to our chains, in this way, part of the internal energy of the chain which is described in terms of its configuration and which is equivalent to Eq. (\ref{eq:hamiltoniano_vara}) is given by the following relation:
%%%%%%%%%%%%%%%%
\begin{equation}
\mathscr{H} \simeq H =\frac{k}{b}\sum_{i,j=1}^{N}\varepsilon_{ij}p_{i},
\label{eq:energia_desvio}
\end{equation}
%%%%%%%%%%%%%%%%
where the weight function $\varepsilon_{ij}$ can take values of $(1)$ or $(-1)$ depending on whether or not the direction of the $i-$th step changes as compared to the previous step (see Fig. \ref{figura3}) and $p_{i}$ represents the probability that each step will find any of its accessible microstates (Eq. (\ref{eq:trueSAW})), $k$ is a constant of units of energy times distance and finally, our model adopts the $b=1$ as the length of Kuhn.

%%%%%%%%%%%%%%%%
%%%%%%%%%%%%%%%%

The bending energy, in its discrete version $H$, describes the behavior of the polymer chain from its tangencial correlations, for example, for a highly correlated polymer chain (Fig. \ref{figura3}, left) this energy will be purely positive (high free energy) what is expected from the Hamiltonian described by the rod and the correlation given in Eq. (\ref{eq:correlation}).\\

In our simulations the calculation of bending energy takes into account the term we call function weight as well as the relative probability of each step of the chain, which has the form of Eq. (\ref{eq:trueSAW}). This energy can be positive or negative depending on the winding of the chain.

%========================================================================
%========================================================================
\section{RESULTS AND DISCUSSION}\label{Resultados}

In order to obtain the results, eleven thousand three-dimensional chains was generated from $2400$ attempts ($N'=2400$).

%========================================================================
\subsection{Chain Length, End-to-end Distance and Radius of Gyration}\label{comprimento}

\begin{figure}[h!]
\begin{center}
\includegraphics[width=8.7cm]{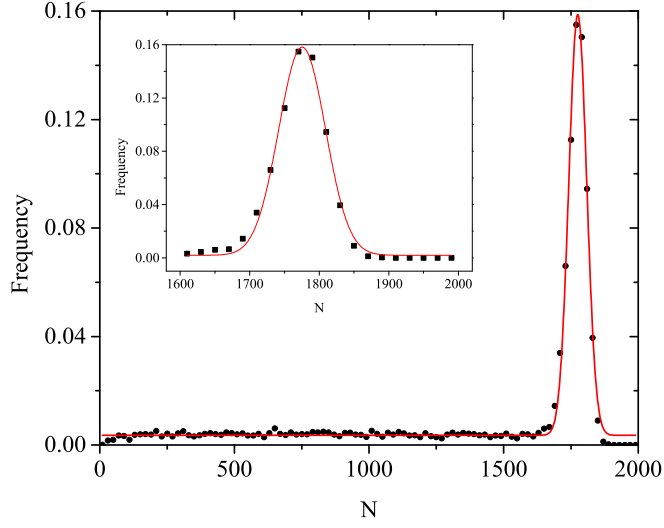}
\caption{Normalized histogram of the number of steps $N$ with  mean $\mu=1775.42$ and standard deviation $\sigma=33.10$. 
Behavior obtained for random chains in $3d$ generated from $N'=2400$. 
The inset illustrates the histogram in a wide interval in which a Gaussian behavior occurs.}
\label{figura4}
\end{center}
\end{figure}

The behavior of the distribution of the number of steps for three-dimensional chains is shown in Fig. \ref{figura4}. 
The maximum of the distribution, is obtained for large chains (compared to $N'$). 
The maximum corresponds to $N = 1775$ steps for the chains generated from $N'=2400$. This way, chains with approximately this number of steps appear with greater probability under the conditions established in the simulation.\\

%%%%%%%%%%%%
%%%%%%%%%%%

\begin{figure}[h!]
\begin{center}
\includegraphics[width=8.7cm]{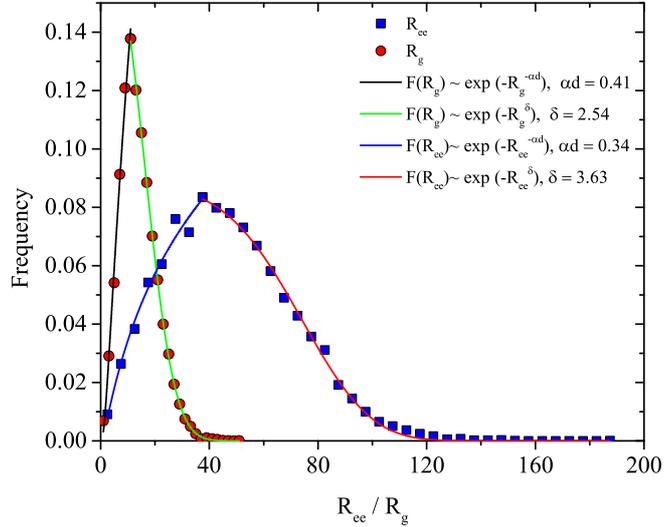}
\caption{Normalized histograms of the radius of gyration ${\bf{R}}_g$ and end-to-end discance ${\bf{R}}_{ee}$. Behavior obtained for the ensemble of random chains generated from $N'=2400$ in $3d$. It is observed that a distributions agrees very well with the expression derived by Lhuillier \cite{Lhuillier_Daniel,Victor_Lhuillier}, both, the ${\bf{R}}_g$ and ${\bf{R}}_{ee}$ distribution.}
\label{figura5}
\end{center}
\end{figure}

\begin{figure}[h!]
\begin{center}
\includegraphics[width=8.7cm]{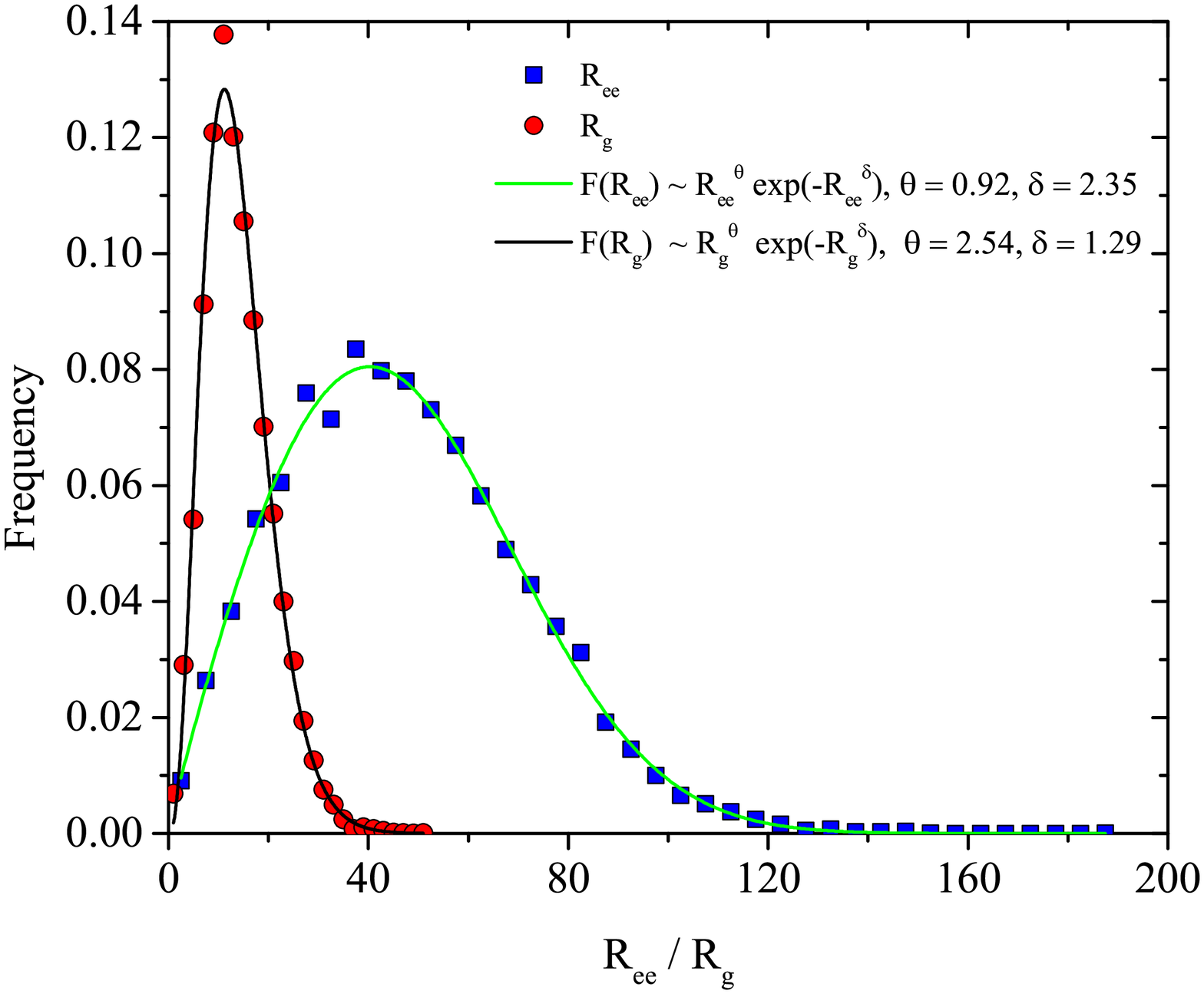}
\caption{Normalized histograms of the radius of gyration ${\bf{R}}_g$ and end-to-end discance ${\bf{R}}_{ee}$. Behavior obtained for the ensemble of random chains generated from $N'=2400$ in $3d$. It is observed that a distributions agrees very well with the function proposed by McKenzie and Moore \cite{McKenzie_1971} and des Cloizeaux  \cite{Cloizeaux_1974,cloizeaux1990polymers}, both, the ${\bf{R}}_g$ and ${\bf{R}}_{ee}$ distribution.}
\label{figura6}
\end{center}
\end{figure}

In Fig. \ref{figura5} and Fig. \ref{figura6}, the distribution of the characteristic distances (${\bf{R}}_{ee}$ and ${\bf{R}}_{g}$) of the chains is shown.  
The graphs in Fig. \ref{figura5} represent adjustments using Lhuillier's proposal \cite{Lhuillier_Daniel}:
%%%%%%%%%%%%%%%%
\begin{equation}
P({\bf{R}}) \sim \exp{(-{\bf{R}}^{-\alpha d}- {\bf{R}}^{\delta})}.
\label{eq:Lhuillier_Daniel}
\end{equation}
%%%%%%%%%%%%%%%%
The distribution behavior can be described separated in two regions that follow different exponential laws. For small values of characteristic distances, the distribution behavior is described by: $\sim\exp(-{\bf{R}}^{-\alpha d})$, where $\alpha=(\nu d-1)^{-1}$ and $d$ is a spatial dimension. For large values of the characteristic distances, the expresion for the distribution is: $~\exp(-{\bf{R}}^{\delta})$, $\delta=(1-\nu)^{-1}$ is the Fisher exponent \cite{Victor_Lhuillier}.

%%%%%%%%%%%%%
%%%%%%%%%%%%%

The Fig. \ref{figura6} shows the graphs of the ${\bf{R}}_{ee}$ and ${\bf{R}}_{g}$ distribution, adjusted according to the Fisher-McKenzie-Moore-des Cloiseaux law \cite{Rubinstein,McKenzie_1971,Cloizeaux_1974,cloizeaux1990polymers}. 
This law is commonly used to describe the distribution of extreme-extreme distance and is of the following type:

%%%%%%%%%%%%%%%%
\begin{equation}
P({\bf{R}}) \sim {\bf{R}^{\theta} \exp({-\bf{R}}^{\delta})}.
\label{eq:Victor_Lhuillier}
\end{equation}
%%%%%%%%%%%%%%%%

The exponent $\theta = (\gamma - 1)/\nu$ characterizes the shorts-distance intra-chain correlations between two segments of a long polymer in a good solvent. The total number of chain conformations is indirectly determined  by the exponent $\gamma$. As opposed to ideal chains, where $\gamma = 1$, for real chains, $\gamma > 1$, i.e. there is a reduction of the probabilities in the distribution of ${\bf{R}}$ for short chains.\\

%%%%%%%%%%%%%
%%%%%%%%%%%%%

It is important to note that while Fisher-McKenzie-Moore-des Cloizeaux theoretical distribution parameters are contructed by fixing the number of steps of the chains and study the distance end-to-end of the different configurations generated, in our approach we generate a random ensemble of chains without fixing a priory the number of steps. However we have been demonstrated that the chains obtained follow the same distribution \cite{Rubinstein,Cloizeaux_1974}.\\

\begin{figure}[h!]
\begin{center}
\includegraphics[width=8.7cm]{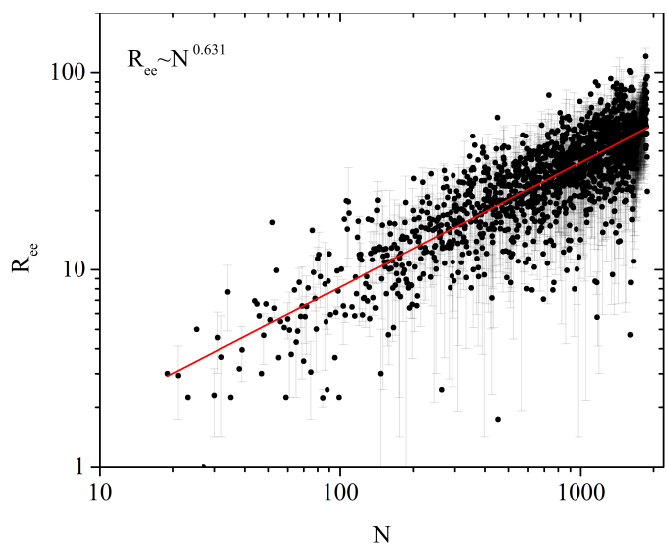}
\caption{Log-log scale representation of the end-to-end distance (${\bf{R}}_{ee}$) as a function of the number of steps ($N$) with its respective value of $\nu$.}
\label{figura7}
\end{center}
\end{figure}

\begin{figure}[h!]
\begin{center}
\includegraphics[width=8.7cm]{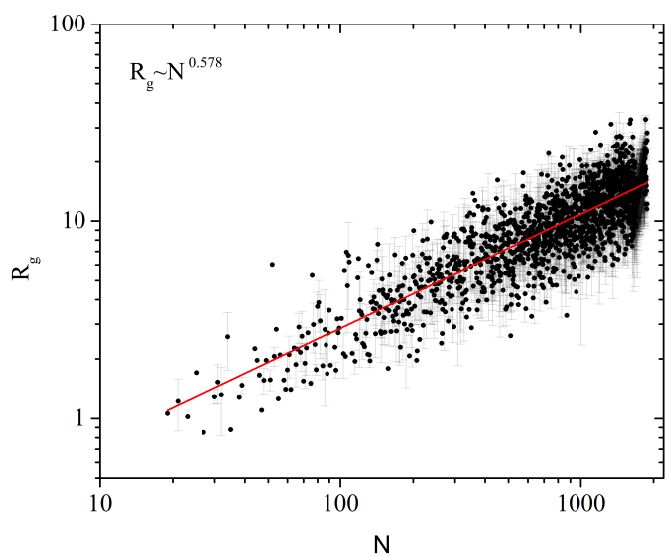}
\caption{Log-log scale representation of the radius of gyration (${\bf{R}}_g$) as a function of the number of steps ($N$) with its respective value of $\nu$.}
\label{figura8}
\end{center}
\end{figure}

The value of Flory exponent $\nu$, which describes the size of the polymer chain, was calculated by computing the mean value of the radius of gyration. The values of $\nu$ obtained from the simulation are incorporated in Fig. \ref{figura7} and Fig. \ref{figura8}.
These results closely approximate the expected theoretical value for the Flory exponent that is $\nu \approx 0.59$ for the three-dimensional case. 
The calculation of $\nu$ from the behavior of ${\bf{R}}_g$ results in a value closer to the expected theoretical value than that calculated from the ${\bf{R}}_{ee}$. \\

The Flory exponent was also calculated indirectly from the delta ($\delta$) exponent that results from the distribution of characteristic distances, shown in Fig. \ref{figura5} and Fig. \ref{figura6}. 
The values of the alpha ($\alpha$) and theta ($\theta$) exponents, the first indirectly and the second directly, were also obtained from these distributions.\\

%%%%%%%%%%%%%%%%
%%%%%%%%%%%%%%%%
%%%%%%%%%%%%%%
%%%%%%%%%%%%%%

The Table \ref{tabla1} shows a comparison between the main critical exponents reported by various authors, both analytically and numerically, and the values reported by our simulations. 
The above results lead us to conclude that although the two behaviors, ${\bf{R}}_{ee}$ and ${\bf{R}}_{g}$, as a function of $N$ fit a power law, the value of the characteristic exponent of ${\bf{R}}_{g}$ is closer to the theoretical value of Flory ($\nu=0.60$ for 3d chains) and to the values reported in \cite{Rubinstein,Ree_dis_polymers,Fluctuating_SoftSphere}($\nu=0.588$).\\

Although both propossed functions fit well the distributions (${\bf{R}}_{ee}$ and ${\bf{R}}_{g}$), the values of the critical exponents obtained for each function, specifically the delta exponent, are different and correspond to different laws. The ${\bf{R}}_{g}$ distribution responds to the law proposed by Lluillier and the distribution of ${\bf{R}}_{ee}$, to the Fisher-McKenzie-Moore-des Cloiseaux law, as already reported in the literature.\\

\renewcommand{\arraystretch}{1.0}
\begin{table}[h!]
\caption{Main critical exponents calculated directly and indirectly from our simulations, compared with values reported by other authors (the results in blue, are values closer to those reported in the literature \cite{Rubinstein, Ree_dis_polymers, Fluctuating_SoftSphere}). Our exponents appear from the fourth row. The fourth and fifth rows show the exponents calculated from the behavior of the characteristic distances as a function of $N$. The exponents calculated using the distribution of ${\bf{R}}_{ee}$ and ${\bf{R}}_{g}$ appear in the next four rows. The exponents of the sixth and seventh rows were obtained using the Lhuillier distribution for both distributions. For the calculation of the exponents that appear in the last two rows, the Fisher-McKenzie-Moore-des-Cloiseuax distribution was used.}
\begin{center}
\begin{tabular}{|l|c|c|c|c|}
\hline
& $\nu$ & $\delta$ & $\alpha$ & $\theta$ \\ 
\hline
Rubinstein \cite{Rubinstein} & $0.588$ & $2.43$ & $1.31$ & $0.28$\\ \hline
Caracciolo et al. \cite{Ree_dis_polymers} & $0.58758\pm0.00007$ & $2.4247\pm0.0004$ & $1.311$ & $0.2680\pm0.0011$ \\
\hline
Vectorel et al. \cite{Fluctuating_SoftSphere} & $0.588$ & $2.38$ & $1.31$    & --- \\
\hline
${\bf{R}}_{g}$ Vs $N$ & \textcolor{blue}{$0.578\pm0.008$} & --- & \textcolor{blue}{$1.36$} & --- \\
\hline
${\bf{R}}_{ee}$ Vs $N$ & $0.631\pm0.011$ & --- & $1.12$ & --- \\
\hline
${\bf{R}}_{g}$ (Lluillier) & \textcolor{blue}{$0.606$} & \textcolor{blue}{$2.54$} & \textcolor{blue}{$1.22$} & --- \\
\hline
${\bf{R}}_{ee}$ (Lluillier)& $0.724$ & $3.63$ & $0.853$ & --- \\
\hline
${\bf{R}}_{g}$ (DC) & $0.225$ & $1.29$ & --- & $2.54$\\
\hline
${\bf{R}}_{ee}$ (DC) & \textcolor{blue}{$0.574$} & \textcolor{blue}{$2.35$} & \textcolor{blue}{$1.295$} & $0.91$\\
\hline
\end{tabular}
\end{center}
\label{tabla1} 
\end{table}

%========================================================================
\subsection{Energy of the Polymeric Chains}\label{energy}

To study the energy of the random polymer chains generated by our simulation, we analyzed the energy given by the interactions between monomer in each chain, as well as the energy associated with the bending of the chains. 
As generated chains are homopolymeric, they are expected to show a uniform behavior or have a low amount of metastable states \cite{entropia_energia,energias}.\\

In the contact potential described in Eq. (\ref{eq:energia_interacao}), the value of the constant $e_{\upsilon_{i}\upsilon_{j}}$ has the information about of interaction energy between the non-continuous and adjacent monomers. This constant adopts the value of  $-1$, for each contact, thus generating a ``folding'' force in the chain, known in proteins as a hydrophobic force \cite{energia_chao_tang}. The sum of all the interactions in chain defines the energy of the system, called \textit{interaction energy} ($E$).\\

To take into account the flexibility of the chain and, consequently, its tangent correlations, we proposed adding to the interaction term $E$, the \textit{bending energy} ($H$) in its discrete version, proposed in Eq. (\ref{eq:energia_desvio}).\\

\begin{figure}[h!]
\begin{center}
\includegraphics[width=7.7cm]{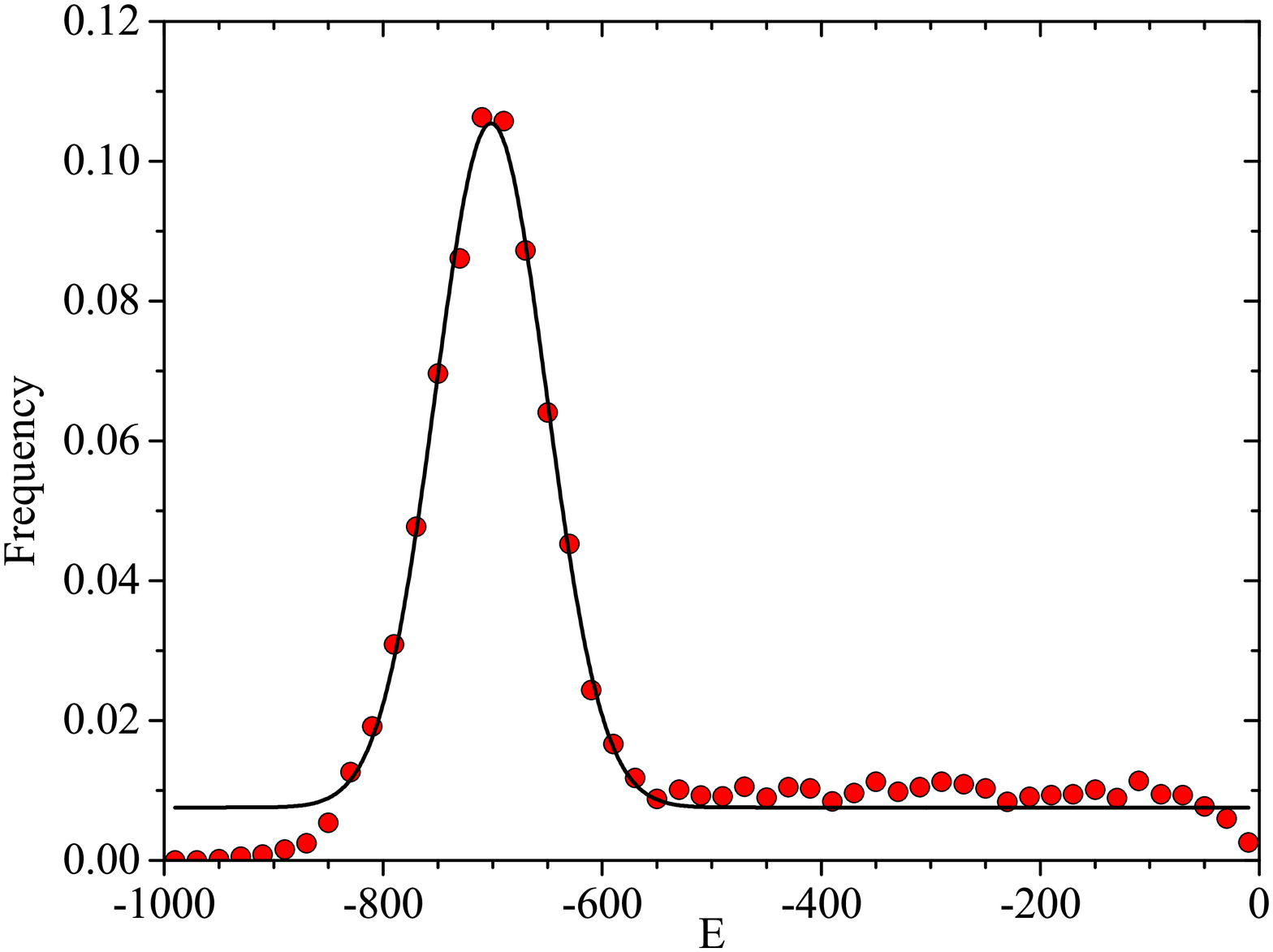}
\includegraphics[width=7.7cm]{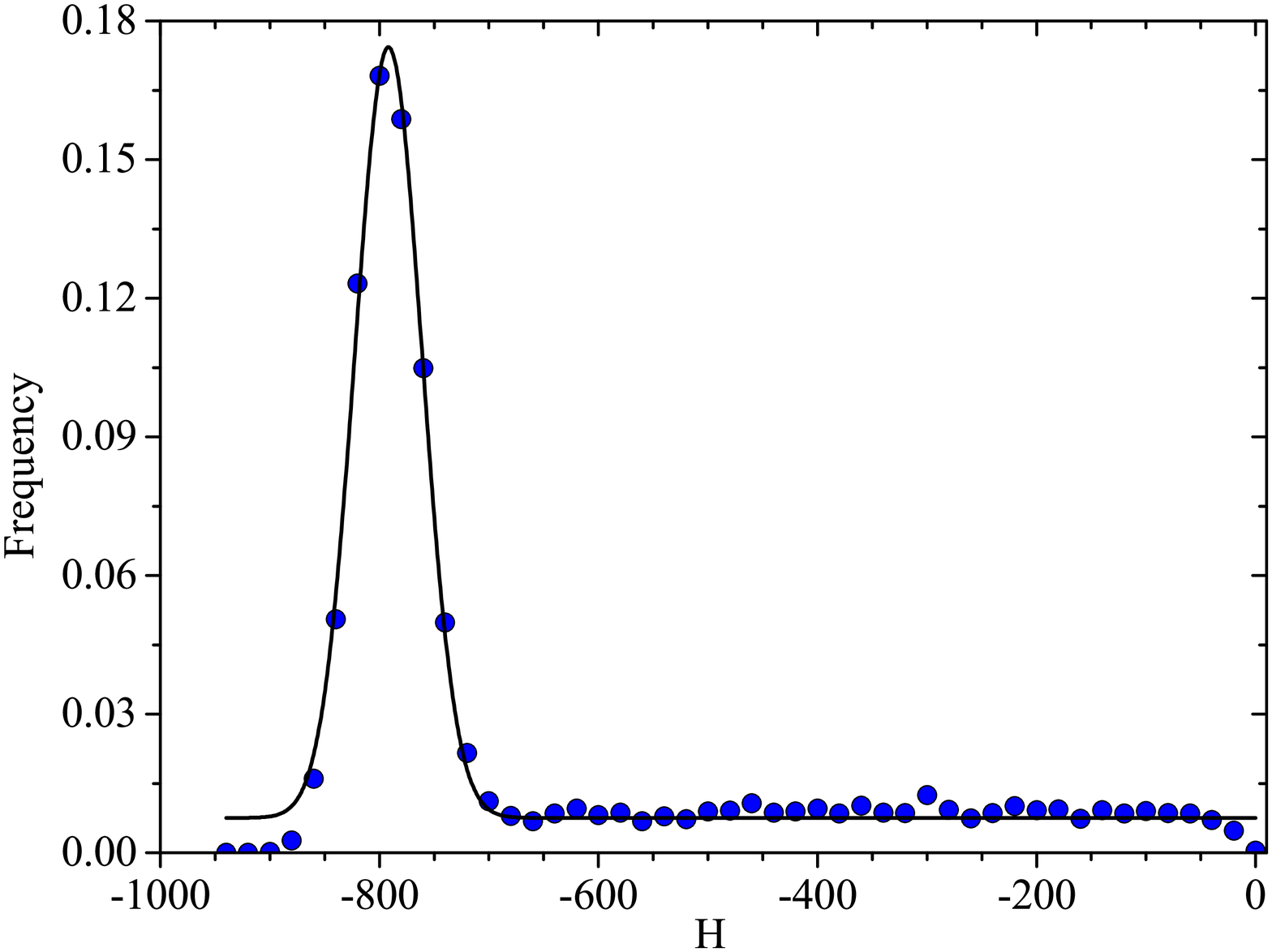}
\caption{Interaction energy $E$ (left side of figure) and bending energy  $H$ (right side) histograms, for three-dimensional chains. Gaussian distributions with standard deviations, $\sigma_{E}=50.66$ and $\sigma_{H}=30.52$, and mean values $\mu_{E}=-701.60$, and $\mu_{H}=-791.77$ respectively.}
\label{figura9}
\end{center}
\end{figure}

In the Fig. \ref{figura9} we can see the characteristic histograms of each energy for three-dimensional chains generated using our algorithm. The shape of the distribution is similar to the results obtained previously for the distribution of number of steps.\\

%%%%%%%%%%%%%
%%%%%%%%%%%%%%%

One of the most important results obtained here is that, even when considering both attractive and repulsive behavior of the chains, the resulting chains have negative energy, because in the SAW models for polymers in a good solvent, the attractions prevail over the repulsions \cite{Rubinstein,polimer_textbook}.\\

\begin{figure}[h!]
\begin{center}
\includegraphics[width=8.7cm]{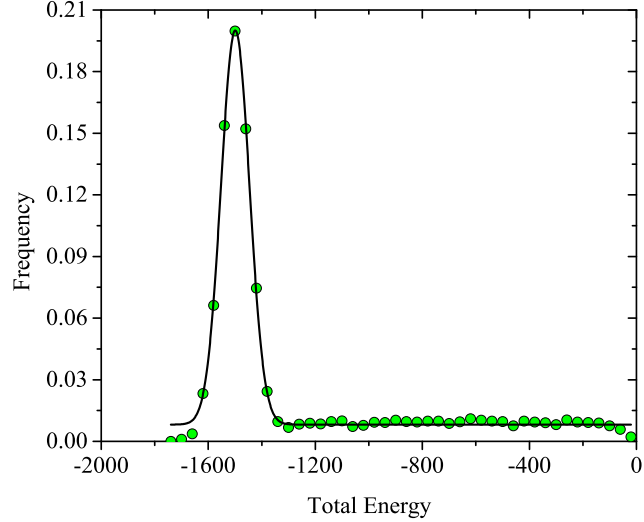}
\caption{Total Energy histogram ($E + H$) for three-dimensional chains with mean with mean $\mu=-1499.14$ and standard deviation $\sigma=53.24$.}
\label{figura10}
\end{center}
\end{figure}

%%%%%%%%%%%%%%%%%
%\begin{figure}[h!]
%\begin{center}
%\includegraphics[width=8.7cm]{Histo_H+E}
%\caption{Total Energy histogram ($E + H$) for three-dimensional chains with mean with mean $\mu=-1499.14$ and standard deviation $\sigma=53.24$.}
%\label{figura10}
%\end{center}
%\end{figure}  
%%%%%%%%%%%%%
The total energy of the chain is described by the sum of interaction and bending energies, which determine the structural configuration.
In the Fig. \ref{figura10}, the total energy distribution is shown. The most probable energy values are  between $-1650$ and $-1350$ energy units, corresponding to long chains. 
The drop in probability density at the tail of the distribution at energy values approaching zero correspond to ``small'' chains.
These small chains can also be identified in the in the probality density function of $N$, which shows the same drop on the left tail of the distribution. 
This may be due to the fact that the total number of configurations, for real chains, is smaller than for ideal chains because of the reduction of probabilities in the distributions of the characteristic distances, mentioned above, for ${\bf{R}}_{ee}\lesssim40$ and ${\bf{R}}_{g}\lesssim10$ (see Fig. \ref{figura5}), which corresponds to $N\lesssim500$.\\

\begin{figure}[h!]
\begin{center}
\includegraphics[width=8.7cm]{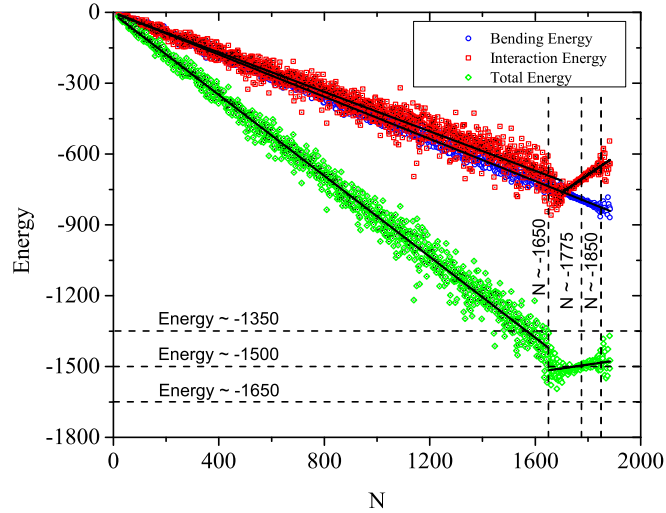}
\caption{Interaction energy $E$ and bending energy $H$ as a function of the number of steps $N$. The horizontal dotted lines represent the values between which the energies are distributed (see Fig. \ref{figura10}). On the other hand, the vertical ones, represent the values between which distribute all sizes of the three-dimensional chains generated.}
\label{figura11}
\end{center}
\end{figure}

Following the same procedure used to study the Flory exponent, the mean values of interaction energy and bending energy were computed for eleven thousand three-dimensional chains. 
The average energy was plotted as a function of N (Fig. \ref{figura11}). 
The plot shows a linear and uniform behavior for the two energy (which is expected \cite{entropia_energia,energias}). When $N$ is large, the behavior of the interaction energy changes from decreasing to increasing, having less negative values, which can be interpreted as representative of more stretched chains, with fewer contacts but with small lengths of persistence still prevailing.\\

%%%%%%%%%%%%%%%%
%%%%%%%%%%%%

%%%%%%%%%%%%
%%%%%%%%%%%%%

The great majority of the three-dimensional chains generated are large ($N>1700$). These chains, with energies between $-1350$ and $-1650$, belong to the second regime of the graph of the total energy and follow a Gaussian distribution (see Fig. \ref{figura10}). The rest of the chains, with energy between $0$ and $-1350$, belong to the first regime of the energy graph and follow a uniform distribution.\\

In the ``discrete'' bending energy approach, the second regime does not appear (see Fig. \ref{figura11}). This term has the same linear behavior, independent of the presence of neighbors interacting with other monomers. The folding of the chains in the large $N$ regime adds negative energy to the system. 
The graph of Fig. \ref{figura11} shows a lower density of points for energies between $-450 and 0$ (corresponding to $0 \lesssim N \lesssim 500$) confirming the decrease in the number of configurations for small chains, as shown in the dot density of Fig. \ref{figura7} and Fig. \ref{figura8} for $N\lesssim 500$.\\

\begin{figure}[h!]
\begin{center}
\includegraphics[width=7.5cm]{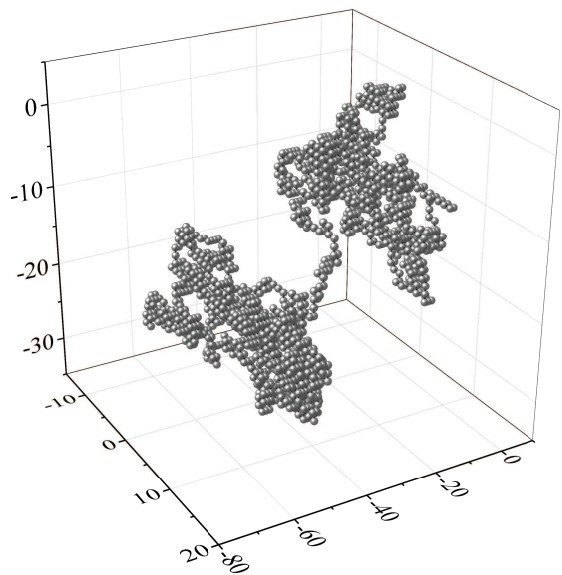}
\caption{Typical structure of $N=1772$ generated by the simulation which shows the formation of two clusters. The spheres represent the monomers of a homopolymeric chain.}
\label{figura12}
\end{center}
\end{figure}

The Fig. \ref{figura12} a typical linear polymeric chain of $N=1772$ generated by the simulation is illustrated. The generated chain forms two clusters separated by a tail. The interaction energy decreases because the chain ``escapes'' and does not have neighbors that contribute to this energy. On the other hand, the bending energy remains constant, since the flexibility of the chain is maintained and causes the chain to continue folding even in the ``bridge'' that separates the clusters.\\

%%%%%%%%%%%%
%%%%%%%%%%%

In the total energy graph of Fig. \ref{figura11}, two regimes can also be seen due to the contribution of the interaction term. For chains of size below $N = 1775$ (mean value of $N$), the energy decreases with $N$ at a rate of $0.86$, practically twice the rate of decrease of each term separately, because the two energy contributes in this regime in the same proportion. For larger chains, the energy increases with $N$ and the characteristic rate of this increase is $0.29$. Here the different contributions of the two types of energy make the increase less significant than in the case of the interaction energy.\\

%========================================================================
%========================================================================
\section{CONCLUSIONS}

In this work, an algorithm based on natural self-avoiding random walk in a cubic network with no boundary was used to generate linear chains. The implementation of this algorithm is simpler and its computational cost is significantly lower than other approaches generally used to generate self-repelling chains. Although our method does not consider all possible configurations for $N$ size of the chain we obtain values of the characteristic critical exponent's of the biopolymers, similar to those reported in literature.  
The distribution of the characteristic distances for the linear chains obtained from our simulation showed a reasonable correspondence with that reported in the literature. Specifically, we prove that although it is possible to fit the two distributions (${\bf{R}}_{g}$ and ${\bf{R}}_{ee}$) using both Lhuillier's theory and the law proposed by Fisher-McKenzie-Moore-des Cloiseaux, the obtained values of critical exponents such as $\delta$, $\alpha$, $\nu$ show us that the distribution of ${\bf{R}}_{g}$ responds better to the theory of Lhuillier, while in the case of the distribution of ${\bf{R}}_{ee}$, the best fit is obtained with the Fisher-McKenzie-Moore-des Cloizeaux function. These results are in full agreement with the literature and reinforce the validity of the algorithm used when characterizing this type of system.\\ 

Both ${\bf{R}}_{ee}$ and ${\bf{R}}_{g}$ resulted in power functions of $N$ and the values of the Flory exponent, in both cases, are quite close to the theoretical value, especially the value corresponding to ${\bf{R}}_{g}$, showing that this characteristic distance is more appropriate when characterizing structurally this type of chains. By comparing the behavior with $N$ of these two distances, we can validate the theoretical linear dependence between these two parameters in the case of real chains.\\

Our study revealed interesting behaviors related to the flexibility of the chains. Small chains are more correlated and consequently, less flexible. Medium-sized chains with N values below $1650$ behave uniformly, being more flexible as $N$ increases.\\

Comparing the behavior of the energy $E$, related to the monomer-monomer interactions of the chains, with the energy $H$, related to its flexibility, we found that chains start with clusters and doubles over distances of the order of ${\bf{R}}_{g}$, increasing its energy. When the chains reach a large number of steps, the interaction energy revealed that the chain stretches and escapes from the cluster, which results in a loss of interaction energy. The behavior of the bending energy reveals that in this escape regime, the chains keep the folding behavior they showed before the escape. This process lasts until the clusters reach a size comparable with the mean radius of gyration.\\ 

The analysis presented in this work, including both the bending and interaction energies is important because it allows differentiating between chains of equal interaction energy but with different structures and hence different bending energies. 
\\

%========================================================================
%========================================================================
\section*{Acknowledgments}
This work has received financial support from CAPES and CNPq (Brazilian Federal Grant Agencies) and from FACEPE (Pernambuco State Grant Agency). We thank Pedro Hugo Figueir\^{e}do, Ra\'{u}l Cruz Hidalgo and Juan Miguel Parra Robles for useful comments of this work.\\

\section*{References}
%% References
%%
%% Following citation commands can be used in the body text:
%% Usage of \cite is as follows:
%%   \cite{key}         ==>>  [#]
%%   \cite[chap. 2]{key} ==>> [#, chap. 2]
%%

%% References with bibTeX database:

\bibliographystyle{elsarticle-num}

\bibliography{bibliograp} 

\end{document}